\documentclass[preprint,12pt]{emulateapj}

\newcommand{\degree}{\mbox{$^{\circ}$}}
\newcommand{\am}{\mbox{\arcmin}}
\newcommand{\as}{\mbox{\arcsec}}

\newcommand\cmv{\mbox{cm$^{-3}$}}

\def\lsim {$\rlap{\raise.4ex\hbox{$<$}}\lower.55ex\hbox{$\sim$}\,$}

\newcommand{\lsun}{\mbox{L$_\odot$}}
\newcommand{\msun}{\mbox{M$_\odot$}}

\newcommand{\lbol}{\mbox{$L_{bol}$}} 

\newcommand{\mean}[1]{\mbox{$\langle#1\rangle$}} 
\newcommand{\av}{\mbox{$A_V$}} 
 
\newcommand{\water}{H$_2$O}

\input{epsf}

\begin{document}

               
\title {The Detection and Characterization of cm Radio Continuum Emission from the Low-mass Protostar L1014-IRS}
\author {Yancy L. Shirley\altaffilmark{1}, 
         Mark J. Claussen\altaffilmark{2},
	 Tyler M. Bourke\altaffilmark{3},
	 Chadwick H. Young\altaffilmark{4},
	 Geoffrey A. Blake\altaffilmark{5}
} 

\altaffiltext{1}{Bart J. Bok Fellow, Steward Observatory, 
University of Arizona, 933 Cherry Ave., Tucson, AZ 85721}
\altaffiltext{2}{NRAO, P.O. Box 0, 1003 Lopezville Road, Socorro, NM 87801}
\altaffiltext{3}{Harvard-Smithsonian Center for Astrophysics, 60 Garden
St. MS42, Cambridge, MA 02138}
\altaffiltext{4}{Nicholls State University, Thibodaux, LA 70310}
\altaffiltext{5}{Division of Geological and Planetary Sciences 150-21,
California Institute of Technology, Pasadena, CA 91125}
 
\begin{abstract}
 
Observations by the Cores to Disk Legacy Team with the 
\textit{Spitzer Space Telescope} 
have identified a low luminosity, mid-infrared source within the dense
core, Lynds 1014, which was previously thought to harbor no internal source.  
Followup near-infrared and submillimeter interferometric observations
have confirmed the protostellar nature of this source by 
detecting scattered light from an outflow cavity and a weak molecular
outflow.  
In this paper, we report the detection of cm continuum emission with the VLA.
The emission is characterized by a quiescent, unresolved 90 $\mu$Jy
6 cm source within $0\farcs 2$ of the \textit{Spitzer}
source.  The spectral index of the quiescent component is
$\alpha = 0.37\pm 0.34$ between 6 cm and 3.6 cm.  A factor of two 
increase in 6 cm emission was detected during one epoch and 
circular polarization was marginally detected
at the $5\sigma$ level with Stokes {\it V/I} $= 48 \pm 16$\% .
We have searched for 22 GHz H$_2$O maser emission toward L1014-IRS, but
no masers were detected during 7 epochs of
observations between June 2004 and December 2006.
L1014-IRS appears to be a low-mass, accreting protostar 
which exhibits cm emission from a thermal jet or a wind, with a variable
non-thermal emission component.  The quiescent cm radio emission is noticeably
above the correlation of 3.6 cm and 6 cm luminosity versus
bolometric luminosity, indicating more radio emission than
expected.  In this paper, we characterize the cm continuum
emission in terms of observations of other low-mass protostars,
including updated correlations of centimeter continuum emission with
bolometric luminosity and outflow force, 
and discuss the implications of recent larger distance estimates
on the physical attributes of the protostar and dense molecular core.
\end{abstract}



\section{Introduction}

It is extremely difficult to identify the incipient stages
of low-mass ($\approx 1$ M$_{sun}$) star formation because 
dense molecular cloud cores obscure nascent protostars. 
Submillimeter dust continuum surveys (e.g., Ward-Thompson et al. 1994,
Shirley et al. 2000, Visser et al. 2002, Kirk et al. 2005)
have identified several dense cores with no \textbf{apparent} internal 
sources, based on the lack of an IRAS point source and the 
diffuse nature of submillimeter dust emission.  
Observations by the Cores to Disk Legacy Team (c2d) with
the Spitzer Space Telescope have identified a few mid-infrared sources
that are embedded near the submillimeter continuum peaks 
of previously classified starless cores (e.g. Young et al. 2004, Bourke et al. 2006).
These new objects are of low-luminosity ($L_{int}$ $\leq 0.1$ \lsun ) 
and presumably low-mass since they were not previously detected 
by IRAS.  Some of these objects may be in the earliest stages of accretion.
These newly identified low-mass, low-luminosity protostars warrant
detailed follow-up studies to determine their evolutionary status.

The first newly identified object detected in the c2d survey, 
L1014-IRS (Young, et al. 2004),  was modeled as a very low-luminosity 
($L_{int} < 0.1$\lsun ), low-mass
($M < 0.1$\msun ) object embedded within the Lynds 1014 dark cloud 
(Lynds 1962) at a distance of approximately 200 pc.
This object has been classified as a VeLLO (Very Low-Luminosity Object)
by the c2d team: an object with an internal protostellar luminosity $\leq 0.1$\lsun\
that is directly associated with a dense molecular core.
The recent study of Morita et al. (2006) has suggested a revised distance
estimate of $400$ to $900$ pc based on the possible age ranges of
nearby T-Tauri stars that are spatially within $2$\degree\  of the L1014
dense core.  However, it is not clear that these T-Tauri stars are directly associated
with L1014.  

Determining the evolutionary state of L1014-IRS has been the subject of 
several follow-up studies. The large scale molecular distribution in the dense 
core was determined by the single-dish mapping survey of Crapsi et al. (2005).  
No evidence for a large scale CO outflow was detected;  however, a molecular outflow was 
detected on small scales with the SMA (Bourke et al. 2005).  Near-infrared observations 
detect scattered light, presumably from the outflow cone, at $1.6$ \micron\ and $2.2$ \micron\ 
(Huard et al. 2006).  The SMA-detected CO outflow is aligned with the direction of the 
near-infrared scattered light cavity.
High resolution molecular observations with BIMA indicate that the protostar
is not at the peak of the molecular and dust column density in the core, but
offset by about $8$\as\ in the plane of the sky (Lai et al., in preparation).
This offset is also seen in (sub)millimeter continuum maps (Young et al. 2004)
and the near-infrared extinction map (Huard et al. 2006).

Despite these significant observational efforts, a single, consistent picture of the
evolutionary state of L1014-IRS has not emerged.
In order to better characterize the physical nature of L1014-IRS, we
have conducted centimeter radio continuum observations using 
five array configurations of the Very Large
Array \footnote{The VLA is operated by NRAO. 
The National Radio Astronomy Observatory is a facility of the 
National Science Foundation operated under a cooperative agreement by 
Associated Universities, Inc.} (VLA).
Centimeter radio continuum emission is well correlated with
the luminosity of protostellar sources (Anglada 1995) and is thought to arise from
shock ionization from protostellar winds (Ghavamian \& Hartigan 1998), 
from interaction of the protostellar jets with dense gas in the 
interface of the outflow cavity (Curiel et al. 1987, 1989, Shang et al. 2004), 
or from accretion shock-driven photoionization (Neufeld \& Hollenbach 1996).

In this paper we report the detection and characterization of the cm radio continuum
emission toward L1014-IRS (\S 3.1).  We compare the detected cm emission with observations
of other low-mass protostars (\S 4.1).  We discuss the non-detections of
22 GHz H$_2$O masers and compare our upper limits to the recent maser monitoring surveys of
low-mass protostars (\S4.2).  We also compare the derived source properties
from the diverse studies of the protostar and dense core in terms of
the range of distance estimates to L1014 (\S 4.3).

\section{VLA Observations}

L1014-IRS was observed during 10 epochs in 5 array configurations (D, A, BnA, B, and C)
with the Very Large Array (Table 1). 
All observations were centered on the published Spitzer mid-infrared source
($\alpha = 21^h 24^m 07^s.51$, $\delta = +49\degree 59\am 09\farcs0$, J2000.0). 
Continuum observations were made at 3.6 cm,
and 6.0 cm, with two polarization pairs at adjacent frequencies, providing a
total equivalent bandwidth of 172 MHz.
We also attempted to detect H$_2$O masers by
observing the J$_{\rm{K}_a \, \rm{K}_c} = 6_{1\, 6} \rightarrow 5_{2\, 3}$
transition at $22.23508$ GHz with, typically, $24.4$ kHz spectral resolution ($0.3$ km/s)
spanning $\pm 20$ km/s velocity coverage.

For the 3.6 and 6.0 cm data, the data were reduced independently 
using the standard routines in \textit{AIPS++} and \textit{AIPS}.  
Complex gain calibration was performed by switching to the nearby quasar 2137+510,
$2.4$\degree\ from L1014-IRS, 
on time-scales of 15 to 30 minutes (Hamaker, Bregman, \& Sault 1996a,b).  
The absolute flux density and bandpass calibration were determined from observations 
of the quasars 3C48 and 3C283.

The Stokes {\it I} and {\it V} images were deconvolved using the 
Cotton-Schwab algorithm (e.g., Schwab 1984) and Clark-H\"ogbom
algorithm (H\"ogbom 1974, Clark 1980) with a few thousand iterations and 
interactive CLEAN regions.  
Imaging the L1014-IRS field was difficult due to the presence of
several bright sources within the VLA primary beam (Figure 1a).  
Special care had to be taken in the CLEANing process (e.g., Cornwell, Braun, \& Briggs 1999)
and multiple reductions with variations in the CLEAN parameters were performed.  We have checked 
the consistency of our images by also reducing the data with the 
standard \textit{AIPS} routines, and the fluxes agree within the statistical errorbars.
Generally, the images are made with natural weighting of the visibilities (Briggs 1995);
however, uniform weighting was used to obtain better angular resolution 
for the full track ($9$ hour) observations on the days of July 1, 2004 and August 21, 2004.

\section{Results}

\subsection{Radio Continuum Detections}

We detected cm radio continuum emission from a source
within $0\farcs 2$ of the Spitzer mid-IR source using the 
VLA at 3.6 cm and 6 cm (Figure 1).  The initial detections were made
during 9 hour tracks at 3.6 cm and 6 cm with the VLA in the D configuration.
Subsequent observations detected the source at 6 cm
in the other three VLA configurations, and again at 3.6 cm in D configuration.  
All of the 6 cm observations,
except for the initial 6 cm detection on August 21, 2004, 
indicate a constant flux density source with an average 6 cm flux of 
$88 \pm 11\, \mu$Jy ($8 \sigma$).
The source is unresolved in all array configurations.  The two
detections at 3.6 cm are also in agreement with each other despite
being separated by 17 months.  The average 3.6 cm flux is $111 \pm 8\, \mu$Jy ($14 \sigma$).
The spectral index between two wavelengths ($\lambda_2 > \lambda_1$) is defined as
\begin{equation}
\alpha = \frac{\ln(S_{\lambda_1}/S_{\lambda_2})}{\ln(\lambda_2/\lambda_1)}\;\; .
\end{equation}
We find an average spectral index between 3.6 cm and 6 cm of
$\alpha = 0.37 \pm 0.34$.

We may estimate the likelihood that the radio continuum emission
is directly related to L1014-IRS and not due to
a background object by calculating the number of background extragalactic
sources with sufficient flux that are expected within our synthesized beam.
The cm continuum source is unresolved in the A configuration observations
at 6 cm with a $0\farcs 46 \times 0\farcs 44$ synthesized beam (natural
weighting).  The position of the 6 cm source ($\alpha = 21^h 24^m 07^s.53$, 
$\delta = +49\degree 59\am 08\farcs 9$, J2000.0) is within $0\farcs 2$ of the
\textit{Spitzer} mid-infrared source.
The 6 cm source is the only source with a flux $\geq 90\, \mu$Jy within
$2^{\prime}$ of the \textit{Spitzer} source.  
The total number of background sources with a flux greater than $S$ $\mu$Jy 
at 6 cm is given by $N(>S) = 0.42(S/30\mu\rm{Jy})^{-1.18}$ per square
arcminute (Fomalont et al. 1991);
therefore, the number of background sources with a flux $\geq 90$ $\mu$Jy
is 0.115 arcmin$^{-2}$.  The probability that a background source lies within our
6 cm synthesized beam is $0.0006\%$.
We believe the centimeter emission is associated with L1014-IRS and not associated with
a background source due to the exceptional positional coincidence and 
strength of the emission.

The cm continuum emission was not constant at all observed epochs.
The 6 cm observations during the epoch of August 21, 2004 
indicate a factor of 2 higher flux.   The 6 cm detection is an
$11\sigma$ detection of $173 \pm 16\, \mu$Jy (Figure 1c).
This indicates that the centimeter flux is variable and may
be an indication of a flare (\S 4.1.2).
We checked the consistency of the August 21 observations
by comparing the 6 cm flux of other unresolved sources within
the L1014 field during the other 6 cm epochs.  No systematic 
increase was detected among the background unresolved
sources on August 21, 2004.

We checked for variability on short timescales by separating and
re-imaging the August 21, 2004 6 cm observations
into four 2 hour time blocks.  The source was found to not vary,
within the uncertainties, during each 2 hour time block.  
This does not rule out a variation
on shorter timescales; however, it is extremely difficult to detect
that variation due to the higher noise level in progressively
shorter time blocks.

A Stokes {\it V} source (circularly polarized) was marginally detected at the 
$5 \sigma$ level at 6 cm on August 21, 2004 
with a flux of $84 \pm 17$ $\mu$Jy at the position of
the 6 cm Stokes {\it I} source (Figure 1d).  
The fraction of circular polarization, $f_c =$ Stokes {\it V/I} $= 48 \pm 16$\%, is quite
high if the detection is significant.  Since the Stokes {\it V} result
was obtained during a full single track of VLA observations,
it will be difficult to confirm until L1014-IRS is observed with
wider bandwidth (e.g., with the eVLA). 
If the circular polarization signal is real, then this observation 
indicates that the radio emission observed on August 21, 2004 must originate  
from a non-thermal mechanism (\S 4.1.2).

\subsection{Water Maser Search}

We searched for \water\ masers during seven epochs spanning 22 months
by observing the J$_{K_a K_c} = 6_{16} \rightarrow 5_{23}$ transition at 22.23508 GHz.
No \water\ masers were detected at any epoch.  The combined $1 \sigma$ 
rms of the non-detection is $3.1$ mJy/beam
with a channel spacing of $24.4$ kHz and a total bandwidth of 3.125 MHz ($\sim$40 km/s).
The individual observations are summarized in Table 1.

\section{Discussion}

\subsection{Centimeter Radio Continuum Emission}

\subsubsection{Steady Component}

Centimeter continuum emission has been detected toward many but not all 
high-mass and low-mass protostars.  The emission is most commonly 
thought to originate from bremsstrahlung
(free-free) emission from ionized gas, although some protostellar 
objects also display non-thermal emission.
For high-mass protostars, the ionization mechanism is photoionization 
usually in the form of
an embedded HII region (Churchwell 1990).  For protostars
that are later in spectral type than B, the ionizing radiation
from the star is not enough to significantly photoionize the surrounding
envelope and an alternative mechanism is needed to explain
the observed emission (Rodr\'iguez et al. 1989, Anglada 1995).  Since nearly 
all low-mass, embedded protostars are
known to have molecular outflows (e.g. Wu et al. 2004), the ionization is postulated to
arise from shocks generated from a jet (e.g., Cohen, Bieging, \& Schwartz 1982, 
Bieging \& Cohen 1989,
Curiel et al. 1987, 1989, Rodr\'iguez \& Reipurth 1996, Shang et al. 2004).  Direct evidence
for this hypothesis comes from observations using interferometers 
at high angular resolution.  Elongated centimeter continuum
emission regions are observed with the same orientation as the large-scale molecular 
outflow toward a few protostars
(e.g., Anglada 1995; Bontemps, Ward-Thompson, \& Andr\'e 1996).  In addition, 
the observed outflow force (\msun\ km/s/yr) theoretically provides
enough energy in the shock to explain the observed centimeter
fluxes toward most low-mass protostars (Cabrit \& Bertout 1992, Skinner et al. 1993,
Anglada 1995).  Finally, the radio spectral index of many protostellar sources
is consistent with optically thin ($\alpha \approx -0.1$) 
to partially optically thick free-free
emission between 3.6 cm and 6.0 cm ($2.0 > \alpha > -0.1$, e.g., Anglada et al. 1998,
Beltr\'an et al. 2001).

The centimeter continuum luminosity (e.g., $L_{3.6} = S_{3.6}D^2$ mJy kpc$^2$)
of low-mass and intermediate-mass protostellar sources
was first cataloged from the literature in the review by Anglada (1995).
Anglada plotted the 3.6 cm luminosity against the bolometric luminosity of
protostars with $L_{bol} < 10^3$ \lsun\ and found a well correlated
relationship ($r = 0.79$), $L_{3.6} = 10^{-2.1} (L_{bol}/1 \lsun )^{0.7}$ mJy kpc$^2$.  
This relationship has formed the basis for predictions of the amount of
centimeter emission that is expected in searches for new low-mass protostars (e.g.,
Harvey et al. 2002, Stamatellos et al. 2007).  
The 3.6 cm luminosity correlation is directly related
to the well established correlation of outflow force vs. bolometric luminosity
(Bontemps et al. 1996, Wu et al. 2004); 
higher luminosity sources drive more
powerful outflows that result in a larger degree of shock ionization and
therefore a larger 3.6 cm luminosity (Anglada 1995).

Since 1995, many more centimeter observations 
have been made and the spectral coverage of the photometry of protostars
has increased.  We have used the
detailed summary tables of Furuya et al. (2003; Table 4) and Anglada (1995) supplemented 
by the surveys of Eiroa et al. (2005) and Anglada et al. (1998) to catalog
the 3.6 cm, 6.0 cm, and bolometric luminosities of detected protostellar sources.
We updated the bolometric luminosity of sources observed in the submillimeter surveys of
Shirley et al. (2000), Mueller et al. (2003), and Young et al (2003).
The resulting sample of $58$ sources at 3.6 cm and $40$ sources at 6.0 cm 
are plotted in Figure 2.  We find updated correlations of
\tiny
\begin{eqnarray}
\log (L_{3.6} / 1 \; \rm{mJy\; kpc}^2) &=& -(2.24 \pm 0.03) + (0.71 \pm 0.01) \log (L_{bol} / 1\; \lsun ) \\
\log (L_{6.0} / 1 \; \rm{mJy\; kpc}^2) &=& -(2.51 \pm 0.03) + (0.87 \pm 0.02) \log (L_{bol} / 1\; \lsun ) \;\; ,
\end{eqnarray}
\normalsize
with correlation coefficients of $r = 0.66$ and $r = 0.74$ respectively.
This sample is not complete as there are many more protostellar
centimeter detections for which no $L_{bol}$ has
been published.  Nevertheless, we have updated the correlation
of Anglada (1995) with twice as many points at 3.6 cm and plotted
the correlation at 6.0 cm for the first time.

For comparison, we have plotted the 3.6 cm and 6.0 cm 
luminosities of L1014-IRS in Figure 2 at the distances of 200, 
400, and 900 pc.  L1014-IRS is above the correlation at all 
distances indicating that we have detected more centimeter
continuum flux than expected.

The spectral index of sources between 3.6 cm and 6.0 cm
is used to argue for the interpretation that the emission mechanism
is consistent with partially optically thick free-free emission.
Optically thin free-free emission is expected to have
a spectral index $\alpha = -0.1$ at centimeter wavelengths.  In
the optically thick limit, $\alpha$ approaches $2.0$ with
intermediate values indicative of partially optically thick
plasmas.  We have also plotted the spectral index of
sources from the literature that have been detected at both
wavelengths and have a published $L_{bol}$ determination
in Figure 2.  No correlation of $\alpha$ is observed 
with bolometric luminosity, probably indicating that the optical depth
associated with the jet's shock ionization is not dependent on
the total protostellar luminosity or the strength of the molecular outflow.
The median spectral index is $\alpha = 0.5$,
with most protostellar sources having flat or positive spectral
indices.  This median value is close to the result expected for
an ionized wind or jet with a $1/r^2$ density gradient (e.g., Panagia \& Felli 1975,
Wright \& Barlow 1975, Reynolds 1986). 
Unfortunately, most of the sources in Figure 2
were not observed at both wavelengths on the same day and
variability may result in significant scatter in the plot.

The spectral index of the quiescent emission of L1014-IRS
($0.37 \pm 0.34$)  is consistent with  
ionized free-free emission with a density gradient.  
The spectral index agrees well with the median of the ensemble 
of protostellar sources measured.

We have also updated the correlation between outflow force, $F_{out}$ (\msun km/s/yr),
and centrimetric luminosity of Anglada (1995) using the molecular outflow compilations
of Bontempts et al. (1996), Furuya et al. (2003), and Wu et al. (2004).  The updated
correlation of 44 sources is weak ($r = 0.55$),
\tiny
\begin{equation}
\log (F_{out} / 1\; \msun \rm{km/s/yr}) = -(3.15 \pm 0.07) + (0.67 \pm 0.03) 
\log (L_{3.6} / 1\; \rm{mJy\;kpc}^2) \;\; .
\end{equation}
\normalsize
This correlation is interpreted as evidence that jets from the
molecular outflow provide enough shock ionization to account to
the observed centimeter continuum emission (e.g., Anglada 1995).
Assuming maximum ionization efficiency, the minimum outflow force needed 
to create the observed level of 3.6 cm flux was estimated by Curiel et al. (1987, 1989)
to be $F_{out} = 10^{-3.5} (L_{3.6} / 1\; \rm{mJy\;kpc}^2)$.  This level is shown as a
dashed line in Figure 2d.  The observed
outflow force toward L1014-IRS ranges from $0.04 - 2.9 \times 10^{-6}$
\msun\ km/s/yr for distances of 200 to 900 pc (see Bourke et al. 2005).
At a distance of 200 pc, the upper limit on the outflow force is a factor
of 2 lower than the Curiel theoretical minimum outflow force.  The upper limit
in the observed outflow force includes estimates of the missing flux due
to interferometric spatial filtering as well as the average opacity corrections
for low-mass protostellar outflows (see Bourke et al. 2005).  Given the
range of uncertainty in these estimates, the observed outflow force is
below, but not necessarily inconsistent with the theoretical minimum outflow force needed
to produce the observed 3.6 cm luminosity.  The disagreement between the observed
outflow force is more pronounced at larger distances, increasing to an order of
magnitude below the theoretical curve for a distance of 900 pc.

The small observed molecular outflow force and the large observed centimeter
continuum luminosities may indicate that another ionization mechanism is operating
in L1014-IRS.   
There are a few
other possibilities that have been discussed in the literature.
We shall analyze the viability of two popular possibilities.

Radiative transfer modeling of emission at $24$ and $70$ \micron\
indicate a flux excess from a disk around L1014-IRS (Young et al. 2004).  
Neufeld \& Hollenbach (1996) postulated that the supersonic infall of
material onto a protostellar disk will create an
accretion shock with enough ionization to generate $\sim 1$ mJy of
continuum emission at centimeter wavelengths at distances of $\approx 200$ pc.  
However, this mechanism does not appear to be able to
provide enough ionization to explain the emission observed toward L1014-IRS.
In order to produce $90$ $\mu$Jy emission at 6 cm at a distance of $200$ pc, 
the protostellar mass has to be $> 2$ \msun\ and the accretion rate onto the disk must
be $> 10^{-4}$ \msun /yr (see Figure 2 of Neufeld \& Hollenbach 1996).  Estimates
of the protostellar mass are very uncertain and highly distance dependent, however,
$2$ \msun\ is likely larger than the L1014-IRS protostar and disk mass, even for the far 
distance estimate
of $900$ pc (see Young et al. 2004).  Furthermore, this accretion rate is an order of
magnitude larger than 
the range of inferred accretion rates from the observed outflow momentum flux and the modeled 
internal luminosity of the source ($\leq 3 \times 10^{-5}$ \msun /yr;
Bourke et al. 2006, Young et al. 2004).  Ionization from an accretion shock does not
appear to be a likely explanation.

A more plausible possibility is that there is a spherical wind
component.  The expected 
continuum emission from self-shocked spherical winds have been modeled by
numerous authors (e.g., Panagia \& Felli 1975, Wright \& Barlow 1975, 
Reynolds 1986, Gonz\'alez \& Cant\'o 2002).  The recent study of
Gonz\'alez \& Cant\'o model a time variable wind that generates internal
shocks (Raga et al. 1990) which produce ionization and centimeter
continuum emission (see Ghavamian \& Hartigan 1998).  Their models
produce centimeter continuum emission of several hundreds of $\mu$-Janskys
and spectral indices that are positive for mass loss rates of $10^{-6}$ \msun /yr
at a distance of 150 pc.  Accounting for the larger distance estimates
of $200$ to $900$ pc and potentially lower mass loss rates 
for L1014-IRS, then this type of emission may still 
account for the flux observed at 3.6 cm and 6 cm ($\sim 100$ $\mu$Jy).
However, the spherical wind models is usually applied to
evolved protostellar sources that are Class II (classical T-Tauri stars) or later
(e.g., Evans et al. 1987).  If L1014-IRS is an older, more
evolved protostar, then this may not be a problem (\S 4.3).

\subsubsection{Variable Component}

While the quiescent component of L1014-IRS is constant over 5 epochs with 
a positive spectral index, the emission properties of L1014-IRS were
significantly different during the single epoch of August 21, 2004.
The 6 cm flux was larger by a factor of two, $S_{6.0} (21 \rm{AUG}2004) = 173 \pm 16$ $\mu$Jy.
Unfortunately, the spectral index of elevated emission was not determined
since L1014-IRS was observed at a single wavelength. Circular
polarization was detected at the $5 \sigma$ level indicating
non-thermal emission.  Thus, L1014-IRS has variable centimeter emission, although the timescale
of the variability is not constrained since it was seen to vary during
only a single epoch.

In general variability of centimeter continuum sources has not been properly
addressed since observations of sources are limited to a few
epochs. There has not been a systematic, monthly monitoring campaign of deeply embedded
sources to characterize their centimeter variability; however, there is observational 
evidence for variability among deeply embedded protostellar sources.  For instance, the 
Class 0 source, B335, is known to vary between an upper limit of  $< 80$ $\mu$Jy
(1994 December) and 
390 $\mu$Jy (2001 January) at 3.6 cm (Avila et al. 2001, Reipurth et al. 2002).
Another example is the variability and purported evidence for jet precession of the
centimeter continuum sources toward IRAS16293 (Chandler et al. 2005).
Variability contributes scatter (de-correlation) of the luminosity correlations
shown in Figure 2.  In additional to the known variable thermal sources, 
several non-thermal protostellar sources are known to be highly variable (e.g. T Tauri stellar
flares, see White 1996); but, most of those objects are more evolved than
deeply embedded protostars.

The observed increase in emission on August 21, 2004 and $5 \sigma$ Stokes {\it V}
detection is indicative of variable non-thermal emission toward L1014-IRS.
While there have been a few high-mass protostars with observed 
negative spectral indices (e.g., Reid et al. 1995; Garay et al. 1996), 
\textbf{most} embedded ($\leq$ Class I) low-mass protostars 
with cm radio emission have positive spectral indices (see Figure 2). 
There are only a few embedded low-mass protostars toward which negative
spectral index,  non-thermal emission is detected (e.g., Shepherd \& Kurtz 1999,
Girart et al. 2002).  One case, R CrA IRS5, was detected 
with significant circular polarization (Feigelson et al. 1998).
The authors postulate that the emission is due to gyrosynchrotron emission and 
may originate from 
magnetic reconnection events associated with flares 
(Feigelson et al. 1998).
The physical mechanism for generating the
radio flare is still not well understood (e.g., Basri 2004) and it
is questionable whether it applies to the embedded phase of low-mass protostars
(see below).  A second case, IRAS 19243$+$2350, is a steep spectrum 
non-thermal source ($\alpha = -0.82 \pm 0.04$) 
that is elongated in the direction of its CO outflow (Girart et al. 2002).
Girart et al. postulate that the emission
may originate from a bi-conical synchrotron source tracing the protostellar jet,
similar to observations of the high-mass source W3(OH) (Reid et al. 1995,
Wilner, Reid, \& Menten 1999).  Non-thermal emission may be
present in low-mass protostellar jets, but the level of emission may be
dominated by the thermal, shock-ionized component of the jet.
In the case of L1014-IRS, this non-thermal jet component would have to
be variable.  We shall discuss several possibilities for the origin
of the observed non-thermal emission toward L1014-IRS.

In order to better understand the non-thermal emission mechanism, 
we estimate the brightness temperature of 
the emission to be $T_b = S_{\nu} \lambda^2 d^2 
/ 2 k R_{\rm{emit}}^2 = 0.14 - 2.8 \times 10^6$ K for distances of 200 to 
900 pc and an emitting region
that is $R_{\rm{emit}} = 1$ AU in size.  The brightness temperature is very sensitive 
to the assumed size of the emitting region.  A $R_{\rm{emit}}$ of 1 AU
is appropriate for a small flare; but, smaller 
when compared to the solar coronal emitting region which is typically less than
$\approx 5$ AU (e.g. Leto et al. 2000).   
Since the emission observed toward L1014-IRS was unresolved,
even in the A-array configuration, then we can only limit the size
of the emission region to $< 90 (D/200 \rm{pc})$ AU.
For instance, if we assumed that the emitting region was $45$ AU (half of our
A-array resolution), then the brightness temperature drops to $< 100$ K.
This is too low even for the steady, thermal free-free component ($T \sim 10^4$ K),
unless the emission was very optically thick.  This cannot be the case
since $\tau >> 1$ would imply $\alpha$ approaching $2.0$ which is not
observed.  While the size of the emitting region is severely unconstrained,
the detection of circular polarization indicates non-thermal emission probably
on small ($<$ few AU) size scales, most likely due to gyrosynchrotron emission
(e.g., Ramaty 1969; Dulk \& Marsh 1982).

Radio flares are observed toward very low-mass objects
including brown dwarfs (Berger et al. 2002, Osten et al. 2006),
late M dwarfs (Berger 2006),
and T-Tauri stars (Bieging \& Cohen 1989; White, Pallavicini, \& Kundu 1992).
The typical level of quiescent emission toward low-mass stars
and brown dwarfs is $100$ $\mu$Jy and the
flares are $1$ mJy (for nearby distances of $\approx 30$ pc)
with significant circular polarization 
($f_c \approx 50$\% ) detected during the flaring events.  
The spectral luminosity at 6 cm of L1014-IRS during the event 
is $L_{\nu } = 4\pi D^2 S_{\nu } = 
4.7 \times 10^{15} (D / 200 \rm{pc})^2$ erg s$^{-1}$ Hz$^{-1}$.  This is about 2400
times larger than the most luminous observed solar flares (Bastian 2004) 
and about $200$ times larger than the flares
observed by Berger (2006) toward late M dwarfs.  If the radio emission
is due to a flare, it must be a powerful flare since L1014-IRS is at least
$20$ times farther away than the average distance of sources detected by
Berger ($\mean{d} = 10.6 \pm 5.1$ pc).  However, it is not larger than
the typical non-thermal flaring emission observed toward young 
T Tauri stars  of $L_{\nu } = 10^{15} - 10^{18}$ erg s$^{-1}$ Hz$^{-1}$ 
(G\"udel 2002).  The ratio of spectral flare luminosity to bolometric luminosity,
$L_{\nu}/L_{bol} = 4 \times 10^{-18}$ Hz$^{-1}$, is also similar to
the ratio observed toward classical T-Tauri stars (see G\"udel 2002).

The timescale over which radio flares toward low-mass stars
are observed tends to occur over minutes to hours.  For instance,
the low-mass brown dwarf, LP944-2, discovered by the 2001 NRAO
summer students (Berger et al. 2002), displays flaring
activity with an average timescale of 10 to 15 minutes.
This is very different from the activity observed toward L1014-IRS
on August 21, 2004.  We detected no evidence for short-term
variability within our detected emission.  The source appears to
have a nearly constant flux that is twice as high as the steady
component for at least an 8 hour period.  This elevated emission then appears
to be longer in duration than that observed during flaring events 
toward low-mass (proto)stars (G\"udel 2002).

An alternative possibility is that the elevated emission is not due to a flare, 
but due to rotational modulation of a non-thermal component 
associated with the magnetic connection between disk and accretion onto the star 
(i.e., Bieging \& Cohen 1989).  Such a mechanism has been postulated for the 
T-Tauri star, V410 Tauri,
with a rotational period of 1.9 days.  The observed emission toward V410 is
$1$ mJy with a negative spectral index.  If the accretion spot is blocked from
view for a fraction of the stellar
rotation period, then it is possibly that we could have observed the source
with the accretion spot is in view on August 21, 2004 and with the accretion spot
blocked from view during the other epochs.  The rotational period of
L1014-IRS must be longer than 8 hours since elevated emission was observed during the
entire 8 hour track.  A negative spectral index was observed toward
V410 Tauri, while a negative spectral has not been observed toward L1014-IRS. 
This hypothesis is highly speculative and would require a regular monitoring campaign 
to test.

Unfortunately, it is currently not possible to strongly constrain the origin
of the non-thermal component toward L1014-IRS.  
The expanded bandwidth of the eVLA is needed to
permit a systematic monitoring campaign with high enough signal-to-noise in only
a few hour observations to routinely check for a Stokes \textit{V} detection and
to determine an instantaneous spectral index.

\subsection{Water Maser Non-detections}

A compact, weak molecular outflow has been detected toward L1014-IRS
(Bourke et al. 2005); therefore it may be possible to detect water
masers if the jet is impinging on dense knots of material
near the protostar.
Previous monitoring of water masers around low-luminosity protostars ($L \leq 10$\lsun )
indicate that the maser activity is highly variable and fugacious 
(e.g., Wilking et al 1994, Claussen et al. 1996, 
Furuya et al. 2003).   There appears to be a threshold
in luminosity ($L_{IRAS} > 25$ \lsun ) above which water masers are always
detected (Wilking et al. 1994).  In general, low-luminosity protostars 
display less water maser activity and have lower isotropic water maser luminosities 
than their high-luminosity counterparts (e.g., Brand et al. 2004, Furuya et al. 2003).
The lowest luminosity protostar with a water maser detection is the Class 0
source, GF9-2, with \lbol\ $= 0.3$ \lsun\ (Furuya et al. 2003).  If
the distance to L1014 is 200 pc, then protostellar luminosity of 
L1014-IRS is a factor of 3 less luminous 
than GF9-2.  A search for water masers toward VeLLOs significantly expands the probed 
luminosity parameter space.

Water maser activity for low-mass protostars is quantified in terms of
the isotropic maser luminosity,  $L_{H_2O}^{iso} = 4 \pi D^2 \int S_{\nu } d\nu$
(e.g. Wouterloot \& Walmsley 1986).
We can calculate the $3 \sigma$ upper limit to the isotropic maser
luminosity of L1014-IRS using the combined rms of the non-detections ($3.1$ mJy/beam in a 
24.4 kHz channel).
The average linewidth of H$_2$O masers sources with $L_{bol} < 10$ \lsun\ is 
$\mean{\Delta v} = 1.2 \pm 0.4$ km/s (Furuya et al. 2003), corresponding to 
3.5 channels in the VLA spectrum.  Therefore, the $3 \sigma$ upper
limit to the maser intensity toward L1014-IRS, assuming the average maser
linewidth, is 11.2 mJy/beam.

Furuya et al. find a correlation between the average isotropic
\water\ maser luminosity and \lbol ; however, there is a mistake in their
conversion from integrated intensity (K km/s) to isotropic water maser
luminosity (\lsun ) due to confusion of kpc with pc (Furuya 2007, private
communication).  The correct conversion equation is given by
\begin{equation}
L_{H_2O}^{iso} = 2.5 \times 10^{-9} \rm{L}_{\odot}\, \left( \frac{D}{200 \rm{pc}} \right)^2
\left( \frac{\int T_A^* dv}{1 \rm{K km/s}} \right) \,\,.  
\end{equation}
We have updated the Furuya et al.
correlation by correcting the conversion factor in their Table 2 and also
adding in points from the literature, namely from a survey of high-mass
star-forming regions and UCHII regions by Wouterloot \& Walmsley (1986),
Palla et al (1993), and a survey of Bok globules by G\'omez et al. (2006).  
Unfortunately, we were not able to add points from some prominent water maser surveys
such as Wilking et al. (1994) and Valderetto et al. (2002) due to incomplete
reporting or differences in the definition of the integrated flux.
The updated correlation for 73 maser sources is shown in Figure 3.  
While at any given epoch, the water maser luminosity of a source 
is highly variable, the mean isotropic water maser luminosity is well-correlated
with \lbol\ ($r = 0.88$) over 6 orders of magnitude in bolometric luminosity:
$L_{H_2O} = 3 \times 10^{-9} \rm{L}_{\odot} \, L_{bol}^{0.94}$.  The relationship is less well
correlated ($r = 0.61$), but does not significantly change, 
if we restict the linear regression to just low-luminosity sources with
\lbol\ $< 100$ \lsun , $L_{H_2O} = 3 \times 10^{-9} \rm{L}_{\odot}\, L_{bol}^{0.93}$.
The best fit slope is near the mean value from the many determinations
made in the literature: e.g., Wouterloot \& Walmsley (1986, $L_{fir}^{0.7}$),
Felli et al. (1992, $L_{fir}^{1.02}$), Palla et al. (1993, $L_{IRAS}^{0.9}$),
Brand et al. (2004, $L_{fir}^{0.81}$).

If we extrapolate this correlation to the internal luminosity of
L1014-IRS at 200 pc ($0.09$\lsun ), then
the expected $L_{H_2O}$ is $3 \times 10^{-10}$ \lsun .  Therefore, our 3$\sigma$ upper limits
($L_{H_2O} < 9.3 \times 10^{-12} (D/ 200 \rm{pc})^2$ \lsun ) are more than an order of magnitude below
the predicted correlation and are significant compared to the strength of other detected
low-luminosity protostars.  The VLA is a very sensitive instrument for water maser
searches, and a systematic monitoring campaign toward newly identified VeLLOs with molecular outflows
is needed to constrain the threshold for the excitation of water masers.

\subsection{The Distance and Evolutionary State of L1014-IRS}

\subsubsection{Distance}

The evolutionary state of L1014-IRS depends on many observational factors
and on the assumed distance to the protostar.  Distances to dark clouds
are very difficult to determine by standard star counting or extinction
techniques (e.g., Schmidt 1975).  For the previously published
papers from the Cores to Disk Legacy group, a distance of $200$ pc was assumed.
Recent observation of T-Tauri stars, presumed to be associated with the L1014 dense
core, indicate that the distance is greater than $400$ pc and
may be as large as $900$ pc (Morita et al. 2006).  The association of the T Tauri
stars with the dense core is not clear as L1014 is located in a region with
multiple dark clouds within a few degrees.  Nevertheless,   
in light of these new results, we discuss the physical properties derived
toward the L1014 core and L1014-IRS with distances that range from $200$ to $900$ pc
(see Table 2).

One potential tool to constrain the distance to L1014
is the use of stellar color excess, $E(B-V)$, of stars in the direction of L1014
(see Bourke et al. 1995).
The MK Classification Extension Catalogue (Morris-Kennedy 1983) 
is used to determine the spectral type and the intrinsic stellar color $(B - V)_0$.  
If the dark clouds near L1014 are associated with each other, then the extinction measured 
toward the MK stars may be used to look for discrete jumps in extinction with distance.
This method is used to produce Figure 4 for stars within $10$\degree\ of L1014.
As evinced by E. E. Barnard's original photographs of this
region (1927), there are many dark clouds within $10$\degree\ of L1014.  One of the
clouds, B164 ($3.8$\degree\ from L1014), is estimated to be $200$ to $240$ pc away
based on its apparent association with the nearby star 80 Cygni (Pagani et al. 1996).
Another cloud, B361 ($3.1$\degree\ from L1014), was estimated to be $350$ pc away (Schmidt 1975).
Unfortunately, many of these clouds
are probably not directly associated with L1014.  As a result, there is no clear jump in
extinction with distance, but instead the jump in extinction
occurs over a range of values from $200$ pc to $500$pc.

Since many of the derived properties have different dependences on distance
in Table 2, we can attempt to use them to limit the range of distances to
the core.  The mass of the dense core has been determined from dust
continuum observations (Young et al. 2004, $M_{d} \propto D^2$) and molecular line observations
(Crapsi et al. 2005, $M_{vir} \propto D$).  The virial mass is equal to the dust mass 
(using the same radius to calculate both quantities) at a distance of $500$ pc; however, 
if we allow for the typical uncertainties in assumed dust opacity, then the
dust mass is usually within a factor of 2 of the virial mass for star-forming
cores (e.g., Shirley et al. 2002, 2003).  Such a range varies the 
distance from $250$ pc to $1000$ pc.  

A better constraint is found from the measured outer radius of the dense core.
The radius was determined by near-infrared extinction
observations of background stars (Huard et al. 2006).  The size
of low-mass star forming cores is limited to typical outer radii of $\leq 60,000$ AU
(Shirley et al. 2002, Jorgensen et al. 2002, Kandori et al. 2005), 
or $\leq 0.3$ pc.  
The measured outer radius is less than $0.3$ pc only for distances of $400$ pc or less.
Therefore, the size of the dense core argues for a closer value of
the distance to L1014.

The current set of observations are unable to place strong constraints on the distance to
L1014, although there is evidence for a distance of $< 500$ pc.  A detailed
spectroscopic study of stars near the periphery of L1014 is needed to better
constrain the distance.

\subsubsection{Evolutionary State}

A fundamental question that remains to be answered is whether L1014-IRS is a newly
formed protostar or a more evolved, accreting object.  In order to address this
question, we must synthesize the current observational information on L1014.
The derived quantities are summarized in Table 2.

First, there is a discrepancy between the modeled density structure of the L1014 dense core
from the near-infrared
extinction map (Huard et al. 2006) and the (sub)millimeter continuum maps
(Young et al. 2004).  The near-infrared data fit a Bonnor-Ebert sphere (Ebert 1955, Bonnor 1956),
a hydrostatic pressure-bounded configuration, 
with a central density that is a factor of 8 higher than the (sub)-millimeter
models.  Both methods have biases that compromise the estimates of the central
density.  The (sub)-millimeter observations detect the core at less than $10 \sigma$,
resulting in a relatively low signal-to-noise radial profile.  
The ability to successfully model the size of the ``plateau''
region of a Bonnor-Ebert profile is marginalized by the poor signal-to-noise and
telescope resolution 
(Shirley \& Jorgensen 2007, in preparation).  Therefore, the (sub)-millimeter models
are a lower limit to the central density of the core.  
The near-infrared
extinction maps are limited by the number of background stars that can be traced
at high extinction (\av\ $> 30$ mag).  In the case of L1014, there are only two stars
above $35$ mag that heavily bias the smoothed density structure (T. Huard 2006, private communication).
Therefore,  near-infrared extinction mapping is unable to reliably determine the size of the 
``plateau'' region for densities above a few $10^5$ \cmv .  The true central
density of L1014 probably lies between $10^5$ \cmv\ and $10^6$ \cmv .  These value
indicate a core that is moderately to strongly centrally condensed, but there are 
dense starless cores that have higher central densities
(e.g., L1544 Evans et al. 2001; L183, Pagani et al. 2003)

The evolutionary state of L1014-IRS may not be directly tied to the density structure of
the core since L1014 may have fragmented in the past.  
There is evidence for this hypothesis since L1014-IRS is
not located at the peak of the dust column density or molecular column density distributions
(Young et al, 2004, Crapsi et al. 2005, Huard et al. 2006, Lai et al. 2007 in preparation),
unlike the case for other Class 0 protostars (see Shirley et al. 2000, Jorgensen et al. 2002).  
The protostar is located 8\as\ to the north of the core peak or $1600$ to $7200$ AU in separation in
the plane of the sky for distances of $200$ to $900$ pc.  
While this distance is within the core's FWHM contour (see J\"orgensen
et al. 2006), it may be evidence of fragmentation.  
Fragmentation of the core
is a more likely possibility due to the higher than average (for low-mass,
dense cores) level of turbulence 
observed toward L1014 (MacLow \& Klessen 2004); it is higher than observed
in most starless cores with a CS linewidth of $0.68$ km/s (Crapsi et al. 2005; Lee, Myers,
\& Plume 2004).  
The protostar may have also moved from its birth-site near the
peak of the core.  This can occur if non-spherical accretion is considered
(Stamatellos et al. 2005).
For a reasonable protostellar velocity of $0.1$ km/s (Walsh et al. 2007),
it would take L1014-IRS $75,000$ to $350,000$ years to move 8\as\ in projection
for distances of $200$ to $900$ pc.  These numbers are within the estimated
lifetimes of the prestellar and Class 0 protostellar phases (Kirk et al. 2005, 
Ward-Thompson et al. 2007).

L1014-IRS has a disk (Young et al. 2004) and is currently accreting material at a relatively low 
rate based on its weak molecular outflow and low luminosity.  The outflow has been active 
for at least several thousand years since an observable outflow cavity has been cleared over a 
region greater than 10\as\ (corresponding to $> 2000 (D / 200 \rm{pc})$ AU) and
has a wide opening angle ($\theta \geq 100$\degree ,Huard et al. 2006).  
L1014-IRS is still classified as a Class 0 source;
however, if the distance is greater than $200$ pc, L1014-IRS
would not be strictly classified as a VeLLO since its internal luminosity is most
likely greater than $0.1$ \lsun .  

The radio continuum observations indicate that L1014-IRS is one of the lowest luminosity
protostars toward which centimeter continuum emission is detected.
The 3.6 cm and 6 cm emission is higher than expected for its protostellar luminosity
and the outflow force is lower than theoretically expected
for its centimeter continuum luminosity.  Shock ionization from the protostellar
outflow is marginally a plausible explanation for nearby distances ($D < 500$ pc).
Another possible explanation may be
a self-shocked spherical wind, although this type of emission is usually
associated with more evolved protostellar objects.  Variable emission is observed and
the properties are consistent with the flaring properties observed toward T-Tauri stars
except for the duration of the elevated emission. The true nature of the variable non-thermal 
component is still a mystery.  The characteristics of both the thermal steady component
and the variable non-thermal component are consistent with centimeter continuum emission
from the later stages of protostellar evolution.
The bulk evidence
seems to indicate that L1014-IRS is not an extremely young protostar (e.g. a first
or second hydrostatic core), but instead seems to have been accreting material
for at least several thousand years to tens of thousands of years and is currently
observed in a low accretion state because it has decoupled from the peak of the
L1014 dense core.

Future observations are needed to better characterize the centimeter
continuum emission variability and to better constrain the mass loss
properties of L1014-IRS.  
Observations with the new eVLA correlator will permit ``instantaneous''
spectral index studies due to the greatly enhanced bandwidth.  Deep
monitoring (rms $< 20$ $\mu$Jy) of the centimeter emission of low-mass protostars will be
possible with integration times less than 1 hour.  If L1014-IRS is
observed during a ``flaring'' event with the eVLA, it will be much easier
to measure the degree of circular polarization and to test for short term
variability.  Near-infrared observations of the protostar with large
aperture telescopes (e.g., Gemini 8-m) 
of accretion diagnostics, such as Br$\gamma$, are needed to
better constrain the accretion properties of L1014-IRS.
Finally, high angular resolution observations of the dense core with
an interferometer with a wide range of $u,v$ coverage and brightness
sensitivity (e.g., eVLA, CARMA, ALMA) will permit detailed analysis of the density
structure of the inner core and the properties of the core in the offset
regions between the protostar and the core column density peak.

\section{Conclusions}

We have detected cm radio continuum emission from L1014-IRS at 3.6 cm
and 6.0 cm.  The emission is characterized by a quiescent, nearly
constant unresolved component of $90$ $\mu$Jy at 6 cm
with a spectral index of $\alpha = 0.37 \pm 0.34$ between 3.6 cm and
6 cm.  We have updated, using recently published observations of
low-mass protostars, the correlations of $3.6$ cm and $6.0$ cm
continuum luminosity versus bolometric luminosity and the correlation of molecular
outflow force with $3.6$ cm luminosity.
The quiescent emission component is consistent with partially ionized
free-free emission, but is above the linear correlation
for low-mass protostars between 3.6 cm luminosity and bolometric
luminosity and below the correlation of outflow force and $3.6$ cm luminosity.  
The quiescent continuum emission may be explained by
shock-ionization from the protostellar jet
or a time-variable wind.  A non-thermal brightening by a factor of 2 of
the 6 cm continuum was detected on August 21, 2004
with $48\pm16\% $ circular polarization at the $5 \sigma$ level.
The true nature of the variable non-thermal component remains a mystery
although the ratio of the flare luminosity to bolometric luminosity is 
consistent with those observed toward T-Tauri flares.
We have also updated the correlation of isotropic water maser luminosity with
bolometric luminosity.
We do not detect water masers toward L1014-IRS, consistent with the highly variable nature 
of water masers around low-luminosity protostars.
Analysis of the derived properties of the L1014 dense core and the L1014-IRS
protostar indicate that it is probably not an extremely young protostar but is
a low-luminosity source that appears to have been accreting for at least several
thousand years and is currently in a low accretion state.  
Properties of the protostar and dense core indicate
it is probably at a distance of $< 500$ pc.

\acknowledgements
We wish to thank Sanjay Bhatnagar, Kumar Golap, and Debra Shepherd 
for their help with \textit{AIPS++}, and are very greatful to
Neal Evans and John Bieging for their suggestions.
Finally, we would like to thank the referee for a very careful reading 
of the manuscript and comments that have greatly improved this paper.




\begin{deluxetable}{lcccccccc}
\tablecolumns{9}
\tabletypesize{\footnotesize}
\tablecaption{VLA Observations\label{tab1}}
\tablewidth{0pt} 
\tablehead{
\colhead{}        & 
\colhead{}        &
\colhead{}        &
\colhead{}	  &
\colhead{}	  &
\colhead{}        &
\multicolumn{3}{c}{L1014-IRS} \\
\colhead{UT Date}        &
\colhead{Config.}        &
\colhead{$\nu$(GHz)}          &
\colhead{Stokes}	 & 
\colhead{Beam\tablenotemark{a}}           &
\colhead{$\Delta \nu$\tablenotemark{b}}   &
\colhead{S$_{\nu}$} & 
\colhead{$\sigma_{S_{\nu}}$}       & 
\colhead{units}                              
}
\startdata 
01 Jul 2004 & D & 8.46 & I  & 5\farcs 9 $\times$ 5\farcs 9   & 172 MHz & 103 & 15  & $\mu$Jy
\\
            & D & 22.23508   & I &        4\farcs 3 $\times$ 3\farcs 4 & 24.4 kHz & ... & 7.6 & mJy
\\
21 Aug 2004 & D & 4.86 & I  & 10\farcs 1 $\times$ 10\farcs 0 & 172 MHz & 173 & 16  & $\mu$Jy
\\
            & D & 4.86 & V  & 10\farcs 1 $\times$ 10\farcs 0 & 172 MHz & 84  & 17  & $\mu$Jy  
\\
            & D & 22.23508   & I &        5\farcs 1 $\times$ 3\farcs 6 & 24.4 kHz & ... & 13 & mJy
\\
21 Nov 2004 & A & 4.86 & I  & 0\farcs 46 $\times$ 0\farcs 44 & 172 MHz & 90  & 18  & $\mu$Jy
\\
11 Jan 2005 & BnA & 22.23508 & I &        0\farcs 12 $\times$ 0\farcs 10 & 24.4 kHz & ... & 9.5 & mJy
\\
07 Mar 2005 & B   & 22.23508 & I &        0\farcs 35 $\times$ 0\farcs 30 & 97.6 kHz & ... & 6 & mJy
\\ 
09 Apr 2005 & B & 4.86 & I  & 1\farcs 7 $\times$ 1\farcs 4   & 172 MHz & 87  & 15  & $\mu$Jy
\\
            & B & 22.23508 & I &          0\farcs 37 $\times$ 0\farcs 30 & 24.4 kHz & ... & 6 & mJy
\\
19 Jul 2005 & C & 4.86 & I  & 5\farcs 0 $\times$ 4\farcs 4   & 172 MHz & 89  & 21  & $\mu$Jy
\\
05 Dec 2005 & D & 8.46 & I  & 9\farcs 5 $\times$ 5\farcs 9   & 172 MHz & 119 & 20  & $\mu$Jy 
\\
29 Nov 2006 & C & 22.23508 & I & 1\farcs 15 $\times$ 0\farcs 97 & 48.8 kHz & ... & 4.9 & mJy \\
28 Dec 2006 & C & 22.23508 & I & 1\farcs 64 $\times$ 0\farcs 83 & 48.8 kHz & ... & 9.8 & mJy \\
\hline
Combined    & ABC & 4.86 & I & 1\farcs 6 $\times$ 1\farcs 5  & 172 MHz & 84  & 10  & $\mu$Jy
\\
\enddata
\tablenotetext{a}{All epochs are imaged with natural weighting except for 01 Jul, 2004 and 21 Aug, 2004
where uniform weighting was used.}
\tablenotetext{b}{The total bandwidth for continuum observations and the channel spacing for H$_2$O maser
observations.}
\end{deluxetable}

\begin{deluxetable}{llccccc}
\tablecolumns{7}
\tabletypesize{\footnotesize}
\tablecaption{L1014 Derived Properties with a Distance Dependence\label{tab2}}
\tablewidth{0pt} 
\tablehead{
\colhead{Property}        & 
\colhead{Dependence}        &
\colhead{D=200pc}        &
\colhead{D=400pc}	  &
\colhead{D=900pc}	  &
\colhead{units}           &
\colhead{Ref.}                        
}
\startdata 
$\Delta\theta_{core}^{pk}$ & $D$ & $\geq 1600$ & $\geq 3200$ & $\geq 7200$ & AU & 1,3,5 \\
L$_{int}$	& $  D^2$ & 0.09  & 0.36  & 1.8	& \lsun\ & 1\\
M$_{env}(R_d)$	& $  D^2$ & 1.7	& 6.8	& 34.4	& \msun\ & 1\\
n$_c^{mm}$	& $  D^{-1/2}$ & 1.5 $\times 10^5$ & 1.1 $\times 10^5$ & 7.1 $\times 10^4$ & \cmv & 1\\
M$_{vir}(R_d)$       & $  D$ & 4.2 & 8.4 & 19.0   & \msun\ & 2\\
R$_o^{nir}$	& $  D$   & 0.13 $-$ 0.16 & 0.26 $-$ 0.32 & 0.59 $-$ 0.72 & pc & 3\\
i$_{out}$	& ... & $> 60$\degree\	& $> 60$\degree\	& $> 60$\degree\	& ...	& 3 \\
R$_{out}$       & $D$   & 540 & 1080 & 2430 & AU & 4 \\
M$_{out}$       & $D^2$ & 0.14 $-$ 1.4 $\times 10^{-4}$ & 0.56 $-$ 5.6 $\times 10^{-4}$ & 0.28 $-$ 2.8 $\times 10^{-3}$ & \msun & 4 \\
t$^{dyn}_{out}$ & $D$   & 700 & 1400 & 3150 & yr & 4 \\
$\dot{\rm{M}}_{out}$ & $D$ & $< 2 \times 10^{-7}$ & $< 4 \times 10^{-7}$ & $< 9 \times 10^{-7}$ & \msun /yr & 4 \\
P$_{out}$       & $D^2$ & 0.24 $-$ 4.5 $\times 10^{-4}$ & 0.1 $-$ 1.8 $\times 10^{-3}$ & 0.5 $-$ 9.1 $\times 10^{-3}$ & \msun\ km/s & 4 \\
E$_{out}$       & $D^2$ & 0.02 $-$ 1.4 $\times 10^{-3}$ & 0.1 $-$ 5.8 $\times 10^{-3}$ & 0.05 $-$ 2.9 $\times 10^{-2}$ & \msun\ km$^2$ / s$^2$ & 4 \\
L$^{mech}_{out}$& $D$   & 0.03 $-$ 2.5 $\times 10^{-4}$ & 0.7 $-$ 5.0 $\times 10^{-4}$ & 0.15 $-$ 1.1 $\times 10^{-3}$ & \lsun & 4 \\
F$_{out}^{obs}$ & $D$   & 0.34 $-$ 6.4 $\times 10^{-7}$ & 0.07 $-$ 1.3 $\times 10^{-6}$ & 0.15 $-$ 2.9 $\times 10^{-6}$ & \msun\ km/s/yr & 4 \\
\hline
L$_{3.6}$       & $D^2$ & 4.4 $\times 10^{-3}$ & $1.8 \times 10^{-2}$  & $9.0 \times 10^{-2}$ & mJy kpc$^2$ & 6 \\
L$_{6.0}$       & $D^2$ & 3.5 $\times 10^{-3}$ & $1.4 \times 10^{-2}$  & $7.1 \times 10^{-2}$ & mJy kpc$^2$ & 6 \\
L$^{flare}_{\nu }$& $D^2$ & 3.9 $\times 10^{15}$ & 1.6 $\times 10^{16}$ & 7.9 $\times 10^{16}$ & erg s$^{-1}$ Hz$^{-1}$ & 6\\ 
L$_{H_2O}^{iso}$& $D^2$ & $< 9 \times 10^{-12}$ & $< 4 \times 10^{-11}$ & $<2 \times 10^{-10}$ & \lsun & 6
\enddata
\tablerefs{1. Young et al. 2004; 2. Crapsi et al. 2005; 3. Huard et al. 2006; 4. Bourke et al. 2006; 5. Lai et al., in preparation, 6. This paper.}
\tablecomments{Source properties: 
$\Delta\theta_{core}^{pk}$ = plane-of-sky distance between L1014-IRS and dense core peak,   
L$_{int}$ = L1014-IRS internal luminosity ($4\pi D^2 \int S_{\nu}^{*} d\nu$),
M$_{env}$ = mass of the envelope, n$_c^{mm}$ = Bonnor-Ebert central density from mm continuum, 
M$_{vir}$ = envelope virial mass, R$_o^{nir}$ = outer radius, 
R$_{out}$ = outflow radius, i$_{out}$ = outflow inclination, 
M$_{out}$ = outflow mass, 
t$^{dyn}_{out}$ = outflow dynamical time, $\dot{\rm{M}}_{out}$ = outflow mass-loss rate,
P$_{out}$ = outflow momentum, 
E$_{out}$ = outflow energy, L$^{mech}_{out}$ = outflow mechanical luminosity, 
F$_{out}^{obs}$ = outflow force, L$_{3.6}$ = 3.6 cm quiescent luminosity ($S_{3.6}D^2$), 
L$_{6}$ = 6 cm quiescent luminosity ($S_{6.0}D^2$), 
L$^{flare}_{\nu }$ = 6 cm flare spectral luminosity ($4\pi D^2 S_{6.0}$), 
L$_{H_2O}^{iso}$ = isotropic H$_2$O maser luminosity ($4\pi D^2 \int S_{\nu} d\nu$).
}
\end{deluxetable}

\clearpage

\begin{figure}
   \vspace*{5in}
   \leavevmode
   \includegraphics{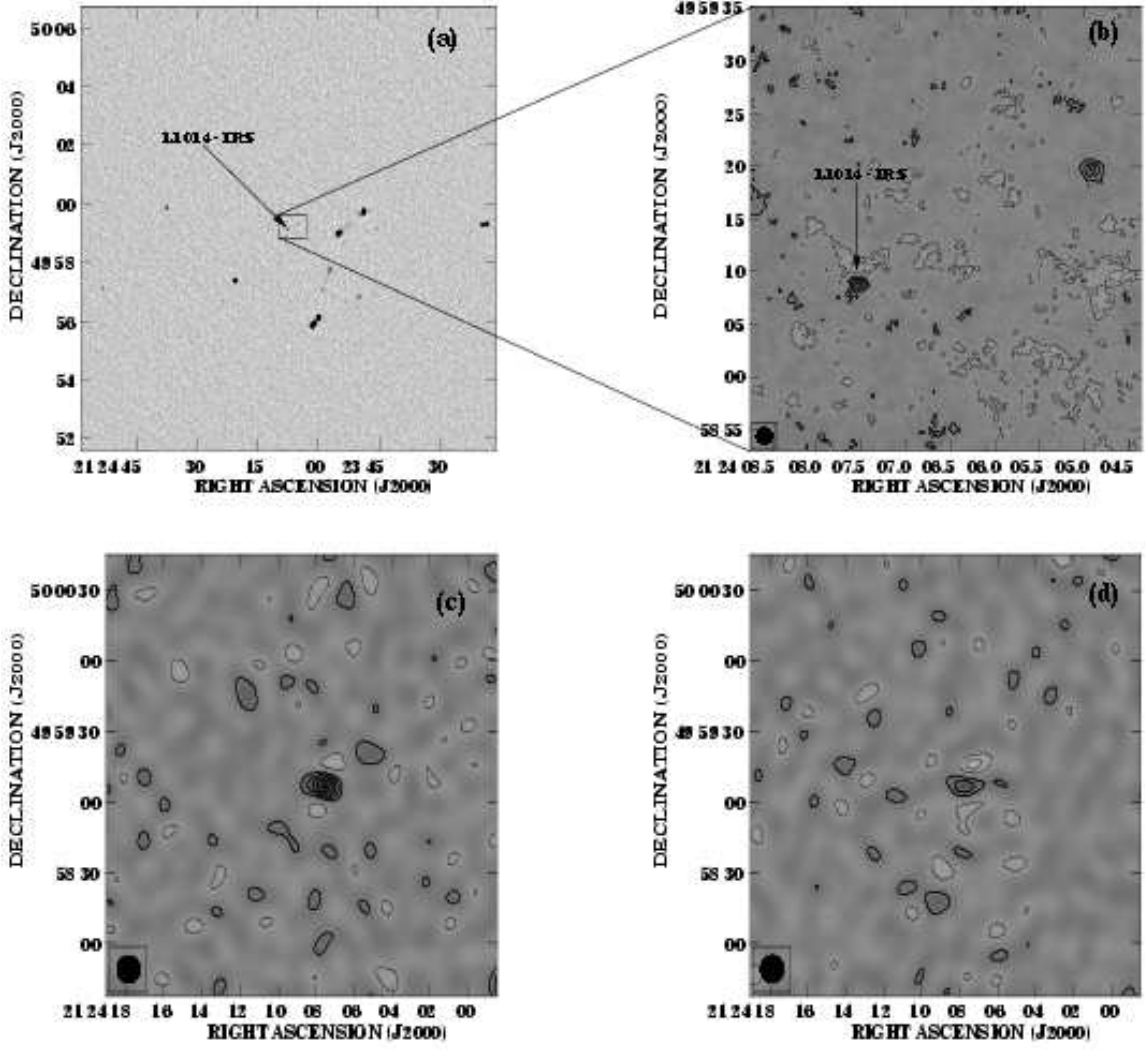}
 \vskip 1.0in
\caption{The initial
detection of L1014-IRS at 3.6 cm (upper left), ABC-array combined 6 cm image (upper right), 
6 cm ``flare'' on 21 August, 2004 (lower left), and Stokes {\it V} 6 cm detection for 21 August, 2004
(lower right). The synthesized beam is shown in the lower left of panels (b) and (c).  
The $2 \sigma$ contours in each panel correspond to: (panel b) 
$\pm 20$ $\mu$Jy/beam;  (panel c) $\pm 33$ $\mu$Jy/beam; (panel d) 
$\pm 33$ $\mu$Jy/beam. Negative contours are shown as light, dash-lines.}
\end{figure}

\begin{figure}
   \vspace*{5in}
   \leavevmode
   \includegraphics{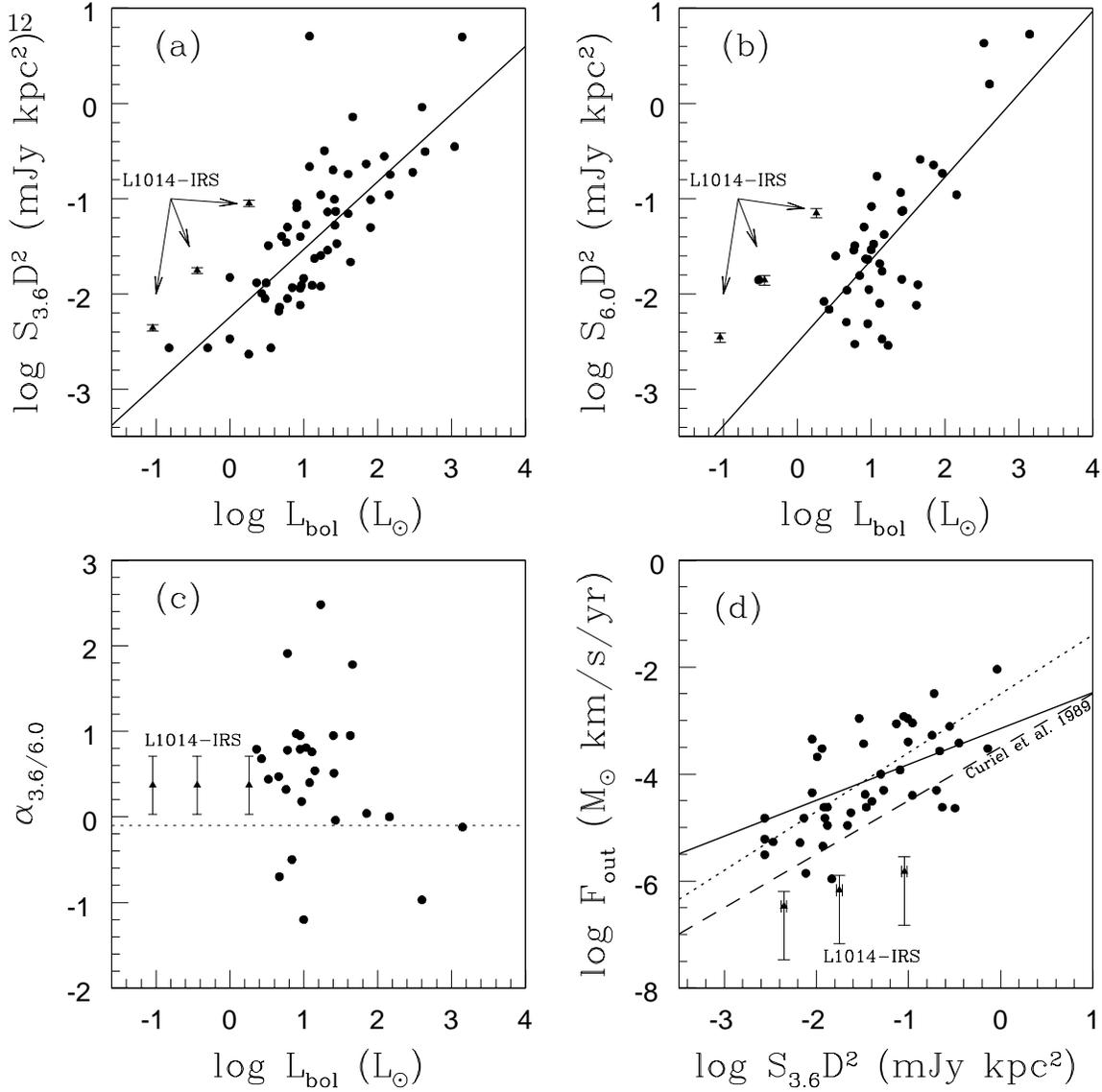}
 \vskip 1.0in
\caption{ Top: Updated correlation of 3.6 cm and 6.0 cm 
luminosity vs. bolometric luminosity for protostars.
The solid line is a linear regression for all points (excluding L1014).
The L1014-IRS 3.6 cm and 6.0 cm quiescent luminosities are shown as trangles for distances of
200, 400, and 900 pc (always increasing left to right in all panels).
Bottom Left:  The spectral index of sources from the literature
with L1014-IRS plotted as trangles at different distances.  The dashed red line
is the $\alpha$ expected for optically thin free-free emission.
Bottom Right: The outflow force ($F_{out} = P_{out}/t_{dyn}$) plotted vs. centimeter continuum 
luminosity.
The solid line is a linear regression to the points (L1014-IRS excluded) while the dotted line
is the Anglada correlation from 1995.  The dashed line is the theoretical minimum relationship
from Curiel et al. (1989).  L1014-IRS is plotted as triangles for the three standard distances.
The errorbars in $F_{out}$ represent the limits determined by Bourke et al. (2005) and are not
statistical errorbars.}
\end{figure}

\begin{figure}
   \vspace*{5in}
   \leavevmode
   \includegraphics{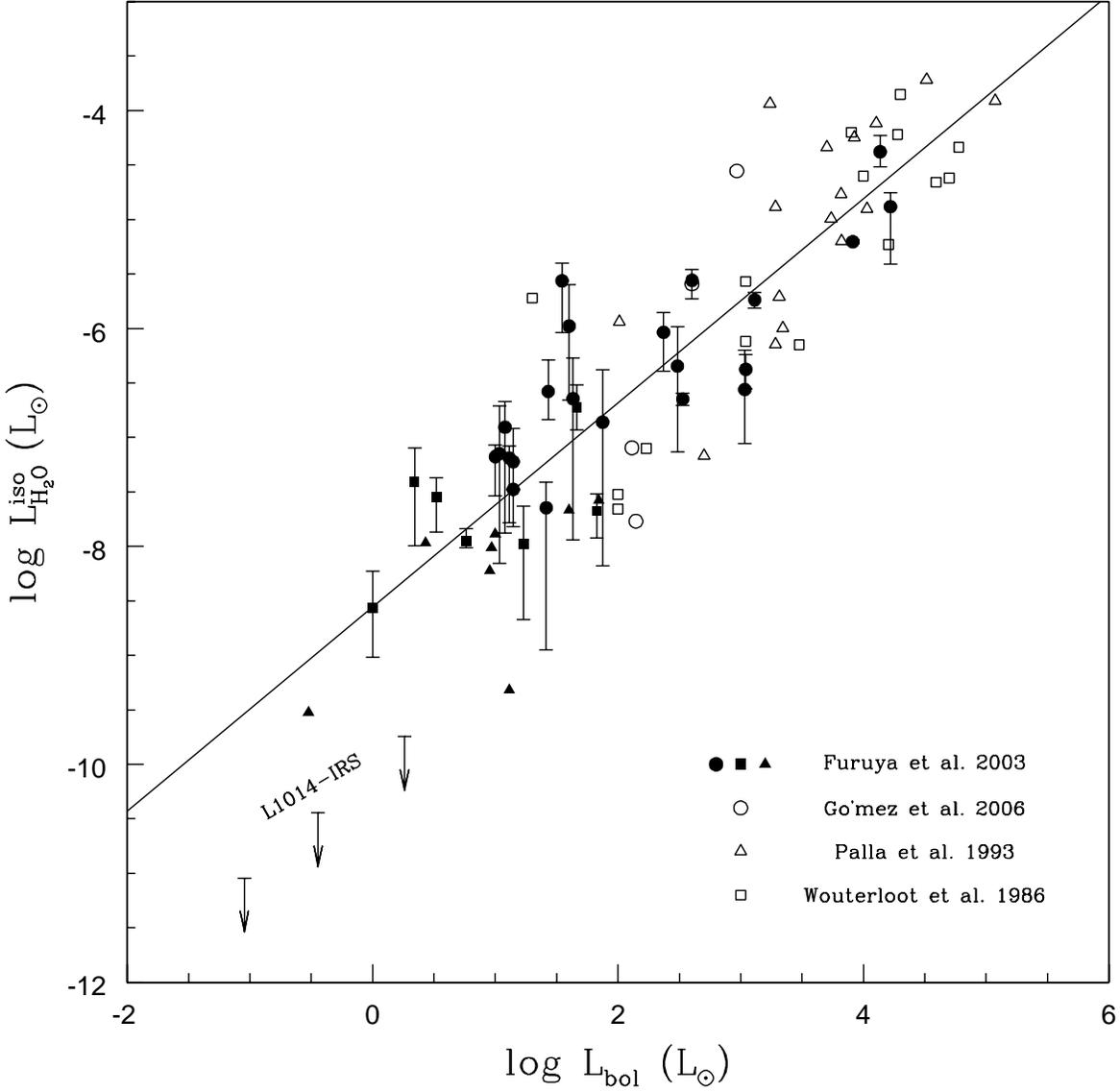}
 \vskip 1.0in
\caption{Updated correlation of isotropic water maser luminosity vs. the
bolometric luminosity.  The filled points are from Furuya et al. (2003),
the open circle points are from G\'omez et al. (2006), the open triangle 
points are from Palla et al. (1993), and the open square points 
are from Wouterloot et al. (1986).
The circular filled points are sources that were detected at all epochs
in the Furuya et al. survey.  The square filled points had at least one
non-detection while the triangle filled points only had one detection.
The errorbars on the Furuya et al. points correspond to the maximal and
minimal isotropic water maser luminosity observed.  The solid line is
the linear regression for all the points in the figure (excluding L1014-IRS).  
The upper limits in the lower left 
are for VLA observations of L1014 at distances of 200, 400, and 900 pc.}
\end{figure}

\begin{figure}
   \centering
   \vspace*{5in}
   \leavevmode
   \includegraphics{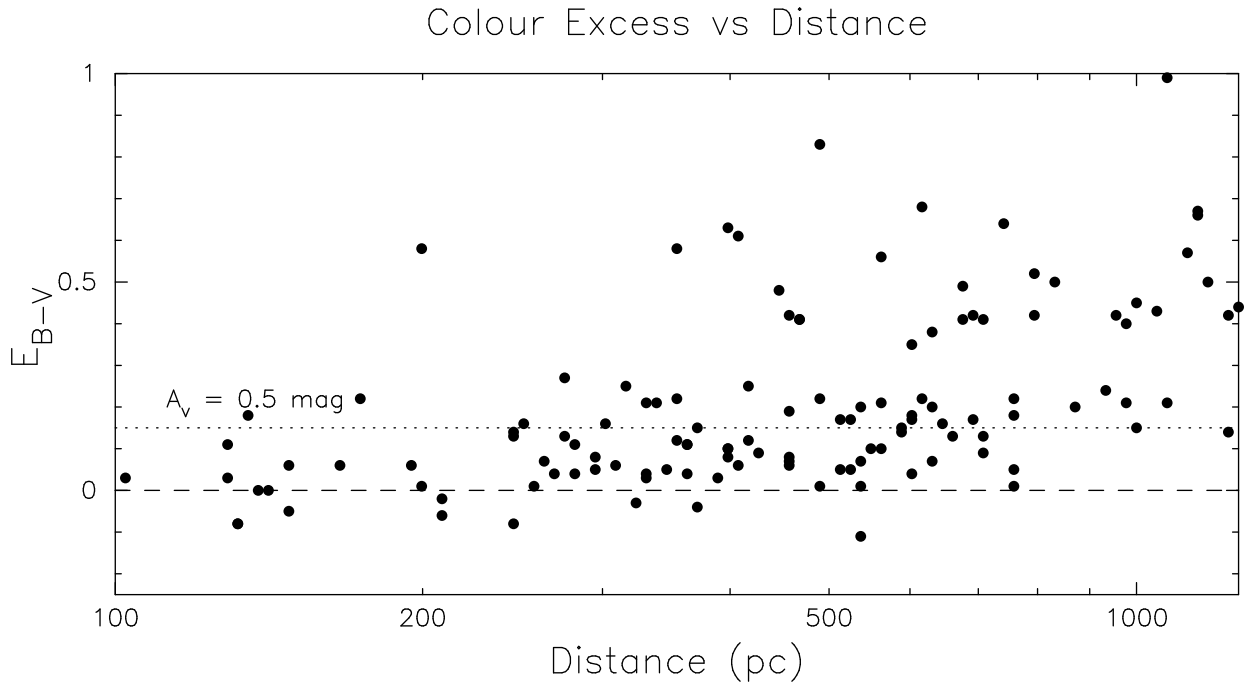}
 \vskip 1.0in
\caption{The color excess vs. distance of MK Extension Catalog stars within 10\degree\ of L1014.
The distance modulus is derived assuming a ratio of selective to total extinction
of $R_V = 3.1 = A_{\rm{V}}/E(B-V)$ (Weingartner \& Draine 2001).}
\end{figure}

\end{document}